\documentclass[doublecol,a4paper,showpacs]{epl2}
\usepackage{graphicx}
\usepackage[all]{xy}
\usepackage{amsmath}
\usepackage{amssymb}
\usepackage{tensor}
\newcommand{\be}{\begin{equation}}
\newcommand{\ee}{\end{equation}}
\newcommand{\ben}{\begin{eqnarray}}
\newcommand{\een}{\end{eqnarray}}
\newcommand{\bes}{\begin{subequations}}
\newcommand{\ees}{\end{subequations}}
\def\bal#1\eal{\begin{align}#1\end{align}}

\newcommand{\bb}{\bibitem}
\newcommand{\sech}{{\rm sech}}
\newcommand{\LL}{{\cal L}}

\begin{document}

\title{Generalized Born-Infeld-like models for kinks and branes}
\author{D. Bazeia\inst{1}\thanks{\email{bazeia@fisica.ufpb.br}} \and M.A. Marques\inst{1}\thanks{{Corresponding author;} \email{mam.matheus@gmail.com}} \and R. Menezes\inst{2,3}\thanks{{\email{rmenezes@dce.ufpb.br}}}}
\shortauthor{D.Bazeia \etal}
\institute{
\inst{1}Departamento de F\'\i sica, Universidade Federal da Para\'\i ba, 58051-970 Jo\~ao Pessoa, PB, Brazil\\ 
\inst{2}Departamento de Ci\^encias Exatas, Universidade Federal da Para\'\i ba, 58297-000 Rio Tinto, PB, Brazil\\
\inst{3}Departamento de F\'\i sica, Universidade Federal de Campina Grande, 58109-970 Campina Grande, PB, Brazil}
\date{\today}

\abstract{In this work we deal with a non-canonical scalar field in the two-dimensional spacetime. We search for a generalized model that is twin of the standard model, supporting the same defect structure with the same energy density. We also study the stability of the defect solution under small fluctuations, which is governed by a Sturm-Liouville equation, and show how to make it stable. The model is then modified and used in the five-dimensional spacetime to construct a thick brane that engenders the first order framework and preserves the twinlike behavior, under tensorial fluctuations of the metric in its gravitational sector.}

\maketitle

\section{1. Introduction}

Defect structures may play important role in structure formation in the early Universe and have been studied in a diversity of scenario involving the cosmic evolution and other issues in high energy physics \cite{vilenkin,manton,weinberg}. Among the several possibilities to build defects, perhaps the best known structures are kinks, vortices, and monopoles. Kinks are the simplest ones, requiring only a single real scalar field. They are static solutions of the equations of motion that connect two neighbor minima of the potential \cite{vachaspati}. Because of their simplicity, kinks can be used to describe specific behavior of several systems in Physics, in particular in magnetic materials, superconductors, topological insulators and in other systems in condensed matter \cite{fradkin}.
 
In this work we focus on kinks and deal with relativistic models described by a single real scalar field in $(1,1)$ spacetime dimensions. We will be concerned with the construction of generalized models \cite{babichev1,gkink1,gkink2,trodden,twinb1,twinb2,twinb3,twinb4,twincol1} of the Born-Infeld type, that support topological structures with the kink profile and energy density as in the corresponding standard model. Such models are called twinlike models \cite{trodden,twinb1,twinb2,twinb3,twinb4,twincol1} because they support the same topological structure, with the same energy density. 

The generalized model that we consider is motivated by several works, some of them dealing with general aspects of defect structures in models with generalized kinematics \cite{babichev1,gkink1,gkink2}, and others with focus on the construction of twinlike models \cite{trodden,twinb1,twinb2,twinb3,twinb4,twincol1}. The model to be investigated is controlled by the parameters $s$ and $M$, and engender non-canonical kinetic terms, with the scalar field being sometimes termed $k-$field. These scalar field theories are similar to those studied in the context of cosmic acceleration \cite{k1,k2}.
Among the generalized models, there is the Born-Infeld concept, originally used to describe nonlinear electrodynamics, with a limiting maximum electric field strength \cite{bi}. Some interesting features appears in this model, such as the absence of birefringence and an electric-magnetic duality; see, e.g., \cite{bi2,bi3} and references therein. Recently, the Born-Infeld idea has been used in different contexts, to investigate how the presence of non-canonical fields may modify the physics in several scenarios, in particular in braneworld models with an extra dimension of infinite extent \cite{bra1,bra2}, in the case of extensions of General Relativity formulated in metric-affine spaces, where metric and connection are regarded as independent degrees of freedom \cite{big1,big2,big3,big4,big5,big6,big7}, in the study concerning holographic entanglement entropy in nonlinear electrodynamics \cite{stri1} and also in the Born-Infeld/gravity correspondence \cite{stri2}. As one knows, the square-root structure of the Born-Infeld modification of gravity introduces a bound in the radicand that contribute to regularize the gravitational dynamics in a way that leads to regular black hole spacetimes and non-singular cosmologies.

We organize the work as follows. In the next Sec.~2 we introduce and investigate the model to explore the requirements to make the model twin to the standard model. We then go on and in Sec.~3 we investigate stability of the solution to show that the model is stable. In Sec.~4 we take the model to construct the corresponding braneworld scenario, and there one shows that it preserves the twinlike behavior. We end the work in Sec.~5, where one includes some comments and conclusions. In this work, we use natural units, $\hbar=c=1$, and this makes the spacetime coordinates to have dimension of inverse of energy. Moreover, we shift the field and the coordinates in a way that make them dimensionless.

\section{2. Model}
\label{model}

Let us start focusing on the Lagrangian density in the two-dimensional $(1,1)$ Minkowski spacetime. We first write the standard density
\bal\label{sta}
{\cal L}_{st}=\frac12\partial_\mu\phi\partial^\mu\phi-U(\phi)=X-U
\eal
where $\phi$ is the scalar field, $X = \frac12\partial_\mu\phi\partial^\mu\phi$ denotes the kinetic term and $U=U(\phi)$ is the potential, in general a function of
the scalar field that can be polynomial or not. We suppose that $U$ has at least the two adjacent minima $v_\pm$. This is the standard model, but here we want to investigate the generalized model, described by the Lagrangian density
\bal
\LL_s &=\frac{M^2}{2(1-s)}\left(\! 1\!-\!\left(1-\frac{2X}{M^2}\right)^s\!\left(\!1+\frac{2U(\phi)}{M^2}\right)^{\!{1-s}}\right)\nonumber\\ \label{lkink}
&\hspace{4mm}+\frac{1-2s}{1-s}X, 
\eal
where $M$ is a mass scale and $s$ is a real positive parameter. The model is controlled by the parameters $s$ and $M$, and engender non-canonical kinetic terms, with the scalar field having a behavior similar to the cases recently studied in \cite{bra1,bra2,big1,big2,big3,big4,big5,stri1,stri2}.

The standard model \eqref{sta} is obtained for $s=0$, and for $s=1$ we get
\be
\LL_{1}=2X-\frac12\left(M^2-2X\right)\ln(P),
\ee
where $P = (M^2+2U)/(M^2-2X)$. We can also take $s=1/2$; in this case one obtains
\be
\LL_{1/2}=M^2-M^2\sqrt{\left(1+\frac{2U}{M^2}\right)\left(1-\frac{2X}{M^2}\right)}.
\ee
This is the model investigated before in \cite{trodden}. It is a Born-Infeld type of model, so it is of current interest in high energy physics.

The model \eqref{lkink} can also be seen as a generalization of the standard case in the sense that, for $M$ very large, the Lagrangian density \eqref{lkink} can be written in the form
\be\label{lasympt}
\LL_{asy}=X-U-\frac{s}{M^2} \left(X+U \right)^2 + \mathcal{O}\left[\frac{1}{M^4}\right],
\ee
which shows that the standard model \eqref{sta} is obtained in the limit $1/M^2\to0$.

In the general case, the equation of motion of the generalized model is given by
\be\label{eomkink}
\partial_\mu\left(\left(\frac{1-2s+sP^{1-s}}{1-s}\right)\partial^\mu\phi\right) + P^{-s}U_\phi=0.
\ee
Here, we are using the notation $U_\phi = dU/d\phi$. The above equation can be rewritten in the form
\be
G^{\alpha\beta}\partial_\alpha\partial_\beta\phi+\left(1+\frac{4sX}{M^2-2X}\right)P^{-s}U_\phi=0,
\ee
where
\be
G^{\alpha\beta} = \left(1-\frac{s(1-P^{1-s})}{1-s}\right)\eta^{\alpha\beta}+ \frac{2sP^{1-s}}{M^2-2X}\partial^\alpha\phi\partial^\beta\phi.
\ee
For $s=1$, the equation of motion is 
\be
\partial_\mu\left(\left(1+\ln(P)\right)\partial^\mu\phi\right) + P^{-1}U_\phi=0.
\ee
For $s$ real and positive, the energy-momentum tensor has the form
\be\label{tmunu}
T_{\mu\nu} = \left(\frac{1-2s+sP^{1-s}}{1-s}\right)\partial_\mu\phi\partial_\nu\phi - \eta_{\mu\nu}\LL.
\ee
The case $s=1$ gives $T_{\mu\nu} = \left(1+\ln(P)\right)\partial_\mu\phi\partial_\nu\phi - \eta_{\mu\nu}\LL$.

Before going further, one has to investigate if the model satisfies the Null Energy Condition (NEC), vital to the stability of the system. The NEC imposes that the energy-momentum tensor \eqref{tmunu} has to obey
\be
T_{\mu\nu}n^\mu n^\nu\geq0,
\ee
where $n_\mu$ is an arbitrary null vector. This condition, for $s\neq1$, leads to
\be\label{nec}
\frac{s(1-P^{1-s})}{1-s}<1,
\ee
and for $s=1$, it leads to $\ln(P)>-1$. As stated before, the potential $U(\phi)$ that appears in Eq.~\eqref{eomkink} presents at least two adjacent minima, $v_\pm$. Then, the simplest solution of the equation of motion \eqref{eomkink} is the uniform field, $\phi_\pm(x)=v_{\pm}$, which leads to $P=1$ and constant energy density. We can also consider static configurations, considering $\phi=\phi(x)$ with boundary conditions $\phi(\pm\infty)=v_\pm$. In this case, we have $X=-{\phi^\prime}^2/2$ and $P = (M^2+2U)/(M^2+{\phi^\prime}^2)$, where the prime stands for the derivative with respect to $x$. Note that, in this case, $P\geq0$ and the equation of motion \eqref{eomkink} reads
\be\label{eomstatickink}
\left(\left(\frac{1-2s+sP^{1-s}}{1-s}\right)\phi^\prime\right)^\prime =P^{-s}U_\phi,
\ee
except for $s=1$, in which it becomes $\left(\left(1+\ln(P)\right)\phi^\prime\right)^\prime =P^{-1}U_\phi$. For the static field, the non-vanishing components of the energy-momentum tensor are
\bes\label{staticemt}
\bal\label{edens}
T_{00}&=\frac{(1-2s){\phi^\prime}^2+(M^2+{\phi^\prime}^2)P^{1-s}-M^2}{2(1-s)}, \\ 
T_{11}&= \frac{(M^2+(1-2s)\,{\phi^\prime}^2)(1-P^{1-s})}{2(1-s)}.
\eal
\ees
For $s=1$, in particular, the components are $T_{00} = {\phi^\prime}^2 +(M^2+{\phi^\prime}^2)\ln(P)/2$ and $T_{11} = ({\phi^\prime}^2-M^2)\ln(P)/2$.

We now set the stressless condition, $T_{11}=0$, to search for stable solutions. By considering the boundary conditions of our problem, $T_{11}=0$ is satisfied by $P=1$. In this case, the NEC in Eq.~\eqref{nec} is satisfied for any $s$ and the equation of motion \eqref{eomstatickink} becomes $\phi^{\prime\prime} = U_{\phi}$, the same equation that one finds for the standard model. It is a second-order differential equation, which can be integrated to give
\be\label{fo}
\frac12{\phi^\prime}^2=U(\phi).
\ee
Therefore, the Lagrangian density \eqref{lkink} and the standard model admits exactly the same stressless solutions. The energy density \eqref{edens} simplifies to
\be\label{rho}
\rho(x) = 2U(\phi(x)),
\ee
as in the standard case. Since the models present the same solutions and energy density, we say that the standard model \eqref{sta} and the generalized model \eqref{lkink} are twinlike models. It is worth commenting that the only requirement for the potential in Eqs.~\eqref{fo} and \eqref{rho} is to support the existence of the kinks connecting its minima. This fact gives a myriad of possibilities. Among them, we have the polynomial class. For instance, if we consider the potential
\be\label{phi4}
U(\phi)=\frac12 (1-\phi^2)^2,
\ee
whose minima are $v_\pm=\pm1$, the Eq.~\eqref{fo} has the solution $\phi(x)=\tanh(x)$ with energy density $\rho(x) = \sech^4(x)$, which can be integrated to give the energy $E=4/3$.

We can follow Ref.~\cite{gkink2} and use an auxiliary function $W=W(\phi)$ such that
\be\label{wdef}
W_\phi = \left(1-\frac{s(1-P^{1-s})}{1-s}\right) \phi^\prime.
\ee
At first glance, the above equation seems to be very complicated. Nevertheless, by using the stressless condition, which imposes $P=1$, we get the first order equation
\be\label{wpflat}
\phi^\prime = W_\phi.
\ee
In this case, the potential is given by
\be
U(\phi) = \frac12 W_\phi^2.
\ee
The above formalism allows to write the energy density as
\be
\rho(x) = \frac{dW}{dx},
\ee
which makes the energy to become $E=|W(v) - W(0)|$. Therefore, it is possible to calculate the energy without knowing the solutions. For instance, for the potential \eqref{phi4}, we have $W(\phi) = \phi-\phi^3/3$. This leads to the energy $E=4/3$, which matches with the value obtained before by a direct integration of the energy density.

\section{3. Stability}
\label{stab}

Let us now focus on the stability of the aforementioned solutions. To do so, we consider time dependent small fluctuations around the static solutions and write the field in the form $\phi(x,t)={\phi}(x)+\sum_i \cos(\omega_i t)\eta_i(x)$, where ${\phi}(x)$ is the solution of Eq.~\eqref{fo}. Plugging this in the time dependent equation of motion \eqref{eomkink} and considering terms up to first order in $\eta_i$, we get the stability equation
\bal
&-\left(\left(1-\frac{4sU}{M^2+2U}\right)\eta_i^\prime\right)^\prime +\left(1-\frac{4sU}{M^2+2U}\right) \nonumber\\ \label{sleq}
&\times\left(U_{\phi\phi}-\frac{4sM^2U_\phi^2}{(M^2+2U)(M^2+2(1-2s)U)} \right)\eta_i =\omega_i^2\eta_i
\eal
This is a Sturm-Liouville eigenvalue equation. 
To study its behavior, it is appropriate to define $A^2 \equiv 1 +4s(1-s)XP^{1-s}/\left((1-2s+sP^{1-s})(M^2-2X)\right)$, which for stressless solutions, $P=1$, becomes 
\be\label{a2}
A^2 = 1-\frac{4sU({\phi}(x))}{M^2+2U({\phi(x)})}.
\ee
We then consider $A^2>0$ to preserve the hyperbolicity of the stability equation \eqref{sleq}. This condition leads to the restriction that $M^2$ has to obey
\be\label{hyperbcond}
M^2 >2\, (2s-1)\, U(\phi(x)), \quad \forall\, x,
\ee
where $\phi(x)$ is the static solution of the model under investigation. The stability equation may be written in the form
\be\label{hstab}
L\eta_i = \omega_i^2\eta_i,
\ee
where $L$ is the Sturm-Liouville operator
\be
L=-\frac{d}{dx}\left(A^2\frac{d}{dx}\right)+ A^2 B,
\ee
where 
\be\label{B}
B=U_{\phi\phi}-\frac{4sM^2U_\phi^2}{(M^2+2U)(M^2+2(1-2s)U)}.
\ee
One can show that the operator $L$ can be factorized as $L=S^\dag S$, with
\bes\label{ss}
\bal
S &= A\left(-\frac{d}{dx} +C -\frac{1}{A}\frac{dA}{dx}\right),\\
S^\dag &= A\left(\frac{d}{dx} +C\right),
\eal
\ees
where $A$ is defined in Eq.~\eqref{a2} and
\be\label{C}
C=\left(1-\frac{4sM^2 U}{\left(M^2+2U\right)\left(M^2+2(1-2s)U\right)}\right)\frac{U_\phi}{\sqrt{2U}}.
\ee
In the case $s=0$, which represents the standard model, we get $A^2=1$, $B=U_{\phi\phi}$ and $C=U_\phi/\sqrt{2U}$. This makes the operators simplify to $S=-d/dx+U_\phi/\sqrt{2U}$ and $S^\dag= d/dx +U_\phi/\sqrt{2U}$.

Therefore, since the operators in Eqs.~\eqref{ss} factorize the stability equation \eqref{hstab}, one can show that 
\be
 \left(S\eta_i\right)^\dag S\eta_i = \eta_i^*\omega_i^2\eta_i \quad\text{or}\quad\left|S\eta_i\right|^2 = \omega_i^2|\eta_i|^2.
\ee
Thus, $\omega_i^2\geq0$ and the model is stable under small fluctuations.

To exemplify, we consider the $\phi^4$ potential in Eq.~\eqref{phi4}. In this case, the Sturm-Liouville equation \eqref{sleq} becomes
\bal
&-\left(\left(1-\frac{2sS^4}{M^2+S^4}\right)\eta_i^\prime\right)^\prime +\Bigg(4-6S^2  \nonumber\\
&+\frac{4sS^4\left(M^2(7S^2-6)+S^4(3S^2-2)\right)}{(M^2+S^4)^2}\Bigg)\eta_i= \omega_i^2\eta_i,
\eal
where $S=\text{sech}(x)$. In this case, from Eqs.~\eqref{a2}, \eqref{B} and \eqref{C} we have
\bes
\bal
A &= \sqrt{1- \frac{2sS^4}{M^2+S^4}},\\
B &= 4-6S^2-\frac{16sM^2S^4T^2}{(M^2+S^4)(M^2+(1-2s)S^4)},\\
C &= 2\left(\frac{2sM^2S^4}{(M^2+S^4)(M^2+(1-2s)S^4)}-1\right)T,
\eal
\ees
with $T=\tanh(x)$. The condition \eqref{hyperbcond} leads to $M^2>(2s-1)S^4$. Since this must be valid for all $x$, in which $0<S\leq1$, we have the condition
\be
M^2 >{2s-1}.
\ee
Since the asymptotic behavior of the Lagrangian density \eqref{lkink} is the standard case, as shown in Eq.~\eqref{lasympt}, we can conclude that, to preserve the hyperbolicity of the stability equation, the model \eqref{lkink} tends to become the standard case as $s$ increases.

We now go back into Eq.~\eqref{sleq} and set $s=0$ to get the stability equation for the standard model \eqref{sta}, which is $-\eta_i^{\prime\prime}+U_{\phi\phi}\eta_i = \omega^2\eta_i$. This shows that the stability of the standard model is studied through a Schr\"odinger-like equation. For a general $s$ in Eq.~\eqref{sleq}, to get an equation similar to the Schr\"odinger equation, we follow the procedure done in Refs.~\cite{gkink1,gkink2} and perform the change of variables 
\be\label{change}
dz=\frac{dx}{A} \quad\text{and}\quad u_i = \eta_i \sqrt{A},
\ee
where $A$ is a function defined by the expression \eqref{a2}. One uses \eqref{change} in order to transform the Sturm-Liouville equation \eqref{sleq} into a Schr\"odinger-like equation
\be
\left[-\frac{d^2}{dz^2}+V(z)\right]u_i = \omega_i^2 u_i,
\ee
where the stability potential is given by

\bal
V(z) &=\left(1- \frac{2s(3M^2+4U)U}{(M^2+2U)^2}\right)U_{\phi\phi} \nonumber\\
     &\hspace{4mm} -\frac{M^2s\left(5M^4\!+\!6(2\!-\!3s)M^2U\!+\!4(1\!-\!2s)U^2 \right)}{(M^2+2U)^3(M^2+2(1-2s)U)}U_\phi^2,
\eal
in which one has to substitute $x$ as a function of $z$ obtained by Eq.~\eqref{change}, a task that is not always possible to do analytically. For $s=0$, we see that $z=x$ and $V(z) = U_{\phi\phi}$, which leads us back to the standard model.

\section{4. Braneworld} \label{branesec}
We now investigate how the model \eqref{lkink} behaves in the five-dimensional curved spacetime, with a warped geometry with a single spatial dimension of an infinite extent. In this case, usually one takes $4\pi G=1$ and considers the action \cite{gubser}
\be\label{sbrane}
S=\int d^4x dy \sqrt{|g|}\left[ -\frac14 R + \LL(X,\phi) \right],
\ee
with $X=\nabla_M\phi\nabla^M\phi/2$ denoting the kinetic term, $y$ standing for the extra dimension and $g$ being the determinant of the metric tensor $g_{MN} = e^{2A}\left( 1,-1,-1,-1, -e^{-2A} \right)$. Here we are using $M,N = 0,1,2,3,4$ and $A=A(y)$ is the warp function. 

In the braneworld context, in order to develop the first order framework we modify the Lagrangian density $\LL_s$ given by Eq.~\eqref{lkink}, adding to it a new term, such that 
\be\label{lbrane}
\LL = \LL_s + \frac13W^2(\phi).
\ee 
One can vary the action \eqref{sbrane} with respect to the metric to get the Einstein equation $G_{MN} = 2T_{MN}$, in which the energy momentum tensor has the form
\be\label{emtbrane}
T_{MN} = \left(1-\frac{s(1-P^{1-s})}{1-s}\right) \partial_M \phi \partial_N\phi - g_{MN}\LL.
\ee
In this section, the limit $s=1$ is similar to the one for the standard case, so we omit it here. For $s$ arbitrary, the equation of motion for the scalar field is given by
\be\label{eombrane}
H^{MN}\nabla_M\nabla_N\phi+\left(1+\frac{4sX}{M^2-2X}\right)P^{-s}U_\phi -\frac23 W W_\phi=0,
\ee
where
\be
H^{MN} = \left(1-\frac{s(1-P^{1-s})}{1-s}\right)g^{MN}+ \frac{2sP^{1-s}}{M^2-2X}\nabla^M\phi\nabla^N\phi.
\ee
As usual, we suppose that the scalar fields only depends on the extra dimension, i.e., $\phi=\phi(y)$. In this case, $X=-{\phi^\prime}^2/2$, with the prime standing for the derivative with respect to $y$. Here, the equation of motion \eqref{eombrane} becomes
\bal
&\left(1-\frac{s(1-P^{1-s})}{1-s}+\frac{4sXP^{1-s}}{M^2-2X}\right)\phi^{\prime\prime} \nonumber \\
&+ 4\left(1-\frac{s(1-P^{1-s})}{1-s}\right)\phi^\prime A^\prime \nonumber \\
&- \left(1+\frac{4sX}{M^2-2X}\right)P^{-s}U_\phi + \frac23 W W_\phi=0.
\eal
The Einstein's equations are
\bes\label{einsteineq}
\bal
A^{\prime\prime} &= \frac{4X}{3}\left(1-\frac{s(1-P^{1-s})}{1-s}\right), \\
{A^\prime}^2 & = \frac19 W^2 + \frac{(M^2-2X(1-2s))(1-P^{1-s})}{6(1-s)}.
\eal
\ees
To get to the first-order framework, we follow Refs.~\cite{gubser,gbrane} and take
\be\label{abrane}
A^\prime = -\frac13 W.
\ee
By substituting this in Eqs.~\eqref{einsteineq}, we get
\bes
\bal
\phi^\prime\left(1-\frac{s(1-P^{1-s})}{1-s}\right) &= \frac12 W_\phi, \\
(M^2-2X(1-2s))(1-P^{1-s}) &= 0.
\eal
\ees
The above equations lead to
\be\label{phibrane}
\phi^\prime = \frac12 W_\phi.
\ee
In this case, the potential $U$ has to have the form $U = W_\phi^2/8$. By comparing the above equation with the flat spacetime equation \eqref{wpflat}, we see that the function $W$ of the braneworld case gives the same solution for the scalar field if $W_{brane}(\phi) = 2 W_{flat}(\phi)$. The first order equations \eqref{abrane} and \eqref{phibrane} for the model \eqref{lbrane} admits the same profiles of the scalar field and the warp function of the standard case, with the Lagrangian
\bal
\LL_{st}=X-\frac18 W_\phi^2 + \frac13 W^2.
\eal
The energy density can be calculated from Eq.~\eqref{emtbrane}; using the first order equations \eqref{abrane} and \eqref{phibrane} one gets
\be\label{rhobrane}
\rho(y) = -\frac32\left(A^\prime e^{2A}\right)^\prime.
\ee
In this case, if $A$ is such that $A^\prime e^{2A}\to 0$ at $y\to\pm\infty$, then the energy of the brane vanishes. Therefore, both the energy density and energy obtained are the same of the standard case. 

As an example, we can consider the $\phi^4$ model of the flat spacetime case, studied in \cite{gubser}, by taking
\be
W_{brane} (\phi) = 2\phi-\frac23\phi^3.
\ee
In this case, the solutions of the first order equations \eqref{abrane} and \eqref{phibrane} are given by
\bal
\phi(y) &= \tanh(y), \\
A(y) &= -\frac19\tanh^2(y) + \frac49 \ln(\sech(y)).
\eal
The energy density \eqref{rhobrane} in this case is
\bal
\rho(y) &= e^{2A}\left(\frac{39}{64} - \frac{31}{48} \tanh(y) -\frac{35}{288}\tanh^2(y)\right. \nonumber \\
&\hspace{4mm} \left.+ \frac{17}{144}\tanh^3(y) - \frac{1}{576} \tanh^4(y) \right),
\eal
and the energy vanishes.

One can investigate the stability of the model under small fluctuations of the field, $\phi\to\phi(y) + \eta(y,x^\mu)$, and of the metric, $g_{\mu\nu} \to e^{2A}(\eta_{\mu\nu}+ h_{\mu\nu})$, with $\mu,\nu=0,1,2$ and $3$, and $g_{44} = -1$ remaining unchanged. If we follow  Ref.~\cite{gbrane}, there are two important conclusions concerning stability of the model \eqref{lbrane} in the braneworld scenario: i) the stability in the gravity sector is controlled by the warp factor; since the model is twin to the standard case, sharing the same scalar field and warp factor profiles, they support the very same stability behavior in this sector; ii) the stability in the scalar
field sector is different because $\LL_{XX}$ does not vanish. In this sense, the thick brane preserves the twinlike behavior, under tensorial fluctuations of the metric in the gravitational sector of the model.\\

\section{5. Ending comments}
\label{end}

In this work we proposed and studied a generalized Born-Infeld-like model that can be seen as twin of the standard model that one usually considers to describe real scalar field in two-dimensional spacetime. The model is controlled by two real and positive parameter $s$ and $M^2$, which may lead to a diversity of new models. 

An interesting thing of the generalized model is that it supports the same kinklike structure and energy density that appear from the standard model.
Also, we considered the linear stability of the solutions, and shown how they are stable against small fluctuations around the static solution.
The model was also modified to construct a thick brane, and there it was shown how it can be used to engender the first order framework and preserve the twinlike behavior in its gravitation sector.

The model is a generalization of the Born-Infeld idea and can be used to investigate other nonlinear effects in curved spacetime, in a way similar to the case studied before in \cite{bra1,bra2,big1,big2,big3,big4,big5,big6,big7,stri1,stri2}. We hope to report on this in the near future.

\acknowledgements{The authors would like to thank the Brazilian agency CNPq for partial financial support. DB thanks support from contracts 455931/2014-3 and 306614/2014-6, MAM thanks support from contract 140735/2015-1, and RM thanks support from contracts 455619/2014-0 and 306826/2015-1.}


\end{document}